# Cu-Modified $SrTiO_3$ Perovskites Toward Enhanced Water-Gas Shift Catalysis: A Combined Experimental and Computational Study

Vitor C. Coletta[a], Renato V. Gonçalves[a], Maria I. B. Bernardi[a], Dorian A. H. Hanaor[b], M. Hussein. N. Assadi[c], Francielle C. F. Marcos[d], Francisco G. E. Nogueira[e], Elisabete M. Assaf[f], Valmor R. Mastelaro[a,*]

[a] São Carlos Institute of Physics, University of São Paulo, SP, 13566-590, São Carlos, SP, Brazil.

[b] Chair of Advanced Ceramic Materials, Technische Universität Berlin, 10623, Germany.

[c] Center for Green Research on Energy and Environmental Materials, National Institute for Materials Science (NIMS), Tsukuba 305-0044, Japan.

[d] Polytechnic School of the University of São Paulo, 05508-010, Department of Chemical Engineering, University of São Paulo, 05508-010, São Paulo, SP, Brazil.

[e] Department of Chemical Engineering, Federal University of São Carlos, 13565-905, São Carlos, SP, Brazil.

[f] São Carlos Institute of Chemistry, University of São Paulo, 13560-970, São Carlos, SP, Brazil.

[*] Corresponding author, email: valmor@ifsc.usp.br

**ABSTRACT:**

The water-gas shift reaction (WGS) is important and widely applied in the production of $H_2$. Cu-modified perovskites are promising catalysts for WGS reactions in hydrogen generation. However, the structure-dependent stability and reaction pathways of such materials remain unclear. Herein, we report catalytically active Cu-modified $SrTiO_3$ (nominally $SrTi_{1-x}Cu_xO_3$) prepared by a modified polymeric precursor method. Microstructural analysis revealed a partially segregated CuO phase in the as-prepared materials. Operando X-ray diffraction and absorption spectroscopy showed the reduction of CuO into a stable metallic phase under conditions of WGS reactions for all compositions. Among the characterized materials, the $x = 0.20$ composition showed the highest turnover frequency, lowest activation energy, and the highest WGS rate at 300 °C. According to density functional calculations, the formation of CuO is energetically less favorable compared with, $SrTiO_3$, explaining why the segregated CuO phase on the $SrTiO_3$ surface is reduced to Cu during the catalytic reaction, while $SrTiO_3$ remains. For $x = 0.20$, the size of the segregated CuO phase is optimum for facilitating the catalytic reaction. In contrast, a higher Cu content ($x = 0.3$) results in an aggregation of smaller CuO particles, resulting in fewer surface active sites and a net decrease in catalytic performance.

# INTRODUCTION

Currently, the most widely employed method for hydrogen production is methane steam reforming, which generates CO as a by-product.[1] In addition to being a harmful gas, carbon monoxide is particularly undesirable for the application of hydrogen fuel in proton exchange membrane fuel cells (PEMFCs), which stand out due to their robustness and versatility but generally employ platinum electrodes, which are contaminated if exposed to CO.[2] To eliminate CO from product gases, the output of methane steam reforming processes generally requires further treatment, which is most often achieved through the water-gas shift (WGS) reaction (CO + $H_2O$ → $CO_2$ + $H_2$) followed by the preferential oxidation (PROX) of CO (2CO + $O_2$ → $2CO_2$) to minimize its concentration and provide further hydrogen output.[1–4] To achieve acceptable conversion rates and reaction kinetics, the WGS reaction is generally carried out in high- and low-temperature stages. In contemporary usage, $Fe_2O_3$-$Cr_2O_3$ and Cu/ZnO-$Al_2O_3$ systems are applied commercially as WGS catalysts for high- and low-temperature reactions, respectively.[1–4]

Alternative copper-loaded catalysts have been proposed for the WGS reaction, notably spinels[5–9] and rare earth oxides.[10–19] Perovskite oxides with $ABO_3$ structures have, in recent years, drawn increasing interest in various catalytic reactions due to their low cost, high stability, and diverse cationic substitutions that can be accommodated in their structure.[20–28] Perovskite strontium titanate ($SrTiO_3$), in a metal-oxide solid-solution or pristine form as support of metal oxides, has been applied in various studies of catalytic reactions such as propane oxidation,[20] methanol decomposition,[21] dry reforming of methane,[22] $NO_x$ reduction,[23] soot combustion,[24] CO oxidation,[25,26] NO reduction,[27] and oxidative coupling of methane.[28] $SrTiO_3$ based materials remain ripe for further exploration towards new catalysts for water-gas shift (WGS) applications.

Studies into $SrTiO_3$ systems containing copper often show that catalytic properties are improved when copper exists as a segregated CuO phase, with the $SrTiO_3$ phase generally acting as a support and CuO as the active phase in many of these catalytic processes.[23–25,27] These studies also show the importance of interactions between the $SrTiO_3$ and CuO phases in relation to catalyst properties. In the case of $SrTi_{1-x}Cu_xO_3$ solid solutions, low catalytic activity is generally observed, as copper atoms were only embedded within the structure of the $SrTiO_3$ phase, while in contrast, the presence of Cu in metallic or oxide forms at the surface of $SrTiO_3$ promotes various catalytic reactions.[23–25,27]

In a previous investigation, we reported the synthesis and structural characterization of $SrTi_{1-x}Cu_xO_3$ ($x$ = 0.03–0.15).[29] It was shown that nominally substitutional copper in the samples segregates apart from the perovskite structure for $x \geq 0.06$ in the form of a secondary CuO phase, which can be reduced to metallic copper at temperatures of ~ 300 °C in a hydrogen atmosphere. Based on these results, we predict that these materials, mainly for the compositions showing CuO segregation, may exhibit useful catalytic functionality. The attributes of the $SrTiO_3$ compound containing Cu atoms concerning different catalysis reactions described previously motivate the further study of this system for WGS applications.

Thus, in the present study, we determined the compositionally dependent efficacy of $SrTi_{1-x}Cu_xO_3$ systems toward the WGS reaction, considering the compositions where segregation of the CuO phase was previously observed.[29] The short- and long-range-order structure, as well as the electronic structure of the catalysts under operando conditions, and the catalytic performance toward the WGS reaction were characterized. A fundamental DFT study was conducted to examine the behavior of the studied materials from first principles to elucidate the aggregation tendencies of Cu atoms both within the host lattice and on the surface of $SrTiO_3$, which eventually leads to the segregation of the CuO phase.

# EXPERIMENTAL AND COMPUTATIONAL SETTINGS

**Catalyst Synthesis.** Cu-modified $SrTiO_3$ nanoparticles were synthesized by a polymeric precursor method (Pechini synthesis), a variant of sol-gel processing that facilitates homogenous cation dispersion through steric entrapment. A more detailed description of such synthesis has been published elsewhere.[29,30] In this synthesis, $SrTi_{1-x}Cu_xO_3$ with $x$ = 0.15, 0.20 and 0.30 was prepared using $Sr(NO_3)_2$, $Ti[OCH(CH_3)_2]_4$ and $Cu(NO_3)_2 \cdot 3H_2O$ as precursors in stoichiometric amounts. Cation chelation was achieved by the addition of citric acid in distilled water with a citric acid:cation molar ratio of 4:1.[29] The citrate chelated solutions were prepared separately for Sr, Cu, and Ti, and the precursor resins were produced by mixing citrate solutions in appropriate quantities and subsequent esterification by the addition of ethylene glycol (EG) with an EG/citric acid mass ratio of 2:3 and evaporation of water under heated stirring. The obtained resins were dried in air at 300 °C for 8 h to obtain precursor powders, which were then heat-treated under $N_2$ flux at 750 °C for 2 h. Samples were further treated with 330 mL/min of oxygen flux at 500 °C for 2 h with a heating rate of 10 °C min$^{-1}$.

**Electron Microscopy.** Scanning transmission electron microscopy (STEM) of the studied samples was performed with an FEI Tecnai G$^2$ F20 microscope operated at 200 kV with a high angle annular dark-field (HAADF) detector. By using this apparatus, EDX mapping images were acquired with an EDAX DX-4 system. For this analysis, particles were dispersed ultrasonically in isopropyl alcohol and deposited on Ni grids.

**X-ray diffraction (XRD).** Conventional *ex-situ* XRD patterns of materials at ambient temperature were recorded using a Rigaku Ultima IV diffractometer with Cu Kα radiation (λ = 1.54 Å) at 2θ = 20°–100° with a 0.02° step size and 3 s per step. Rietveld refinement procedures were carried out using GSAS software.[31] For these refinements, crystallographic information files of $SrTiO_3$ (CIF-80873), CuO (CIF-16025), and $SrCO_3$ (CIF-15195) were used as reference structures.

Operando X-ray diffraction measurements were conducted during representative WGS reactions at the XPD beamline of the National Laboratory of Synchrotron Light (LNLS, Campinas - Brazil, proposal number 17001) using a Huber diffractometer, Arara furnace, and Mythen detector. XRD patterns were collected over the 2θ range 35 to 45°, with radiation of λ = 1.5498 Å using a Si (111) monochromator. Diffraction measurements were performed for samples that had previously been reduced at 250 °C under a 5% vol. $H_2$/He atmosphere for 30 min. After the reduction, the furnace was purged with inert gas, and the temperature was raised to 350 °C. For the investigative WGS measurements and operando XRD, the gas mixture was fixed at 5 vol%. CO/He at a flux rate of 60 mL/min, and water was supplied by a saturator at a molar ratio CO:$H_2O$ equal to 1:3. The feed composition of 5 vol% CO/He was supplied by Brooks Instrument mass flow controllers, while the flow control of He that fed the saturator with water was controlled through a rotameter. The temperature of the saturator was kept at 298 K during the experiment to keep constant the molar ratio CO: $H_2O$ equal to 1:3. The temperature of the saturator was measured with K type thermocouple. Besides that, the temperature of the line from the saturator to the reactor was maintained heating.

**X-ray Absorption Spectroscopy (XAS).** X-ray absorption near-edge spectroscopy (XANES) was performed in operando at the Cu K-edge (8979 eV) in the XAFS2 synchrotron beamline at LNLS (LNLS, Campinas - Brazil, proposal number 18857). Powders were blended with boron nitride and compacted into pellets, which were placed under the normal incidence of the X-ray beam. The parameters for spectral recording and treatment are as previously reported.[29] Similar to operando XRD experiments, the samples were reduced, and XANES spectra were collected during the WGS reaction at 350 °C using a gas mixture of 5 vol%. A saturator supplied CO/He at a flow rate of 60 mL/min and water at a molar ratio CO:H$_2$O equal to 1:3.

**Temperature-Programmed Reduction (TPR).** To evaluate the redox behavior of materials synthesized in the present work, we performed TPR measurements using a Micromeritics Chemisorb 2750 with 100 mg of sample and a 25 mL/min flux rate of 10 vol%. H$_2$/Ar gas. H$_2$ consumption was measured using a thermal conductivity detector (TCD).

**Dispersion of metallic particles.** N$_2$O chemisorption was performed to gauge the surface area and dispersion of metallic copper in reduced materials. For this analysis, 200 mg samples were used. First, the samples were reduced by 3 vol%. H$_2$/Ar mixture at a flux of 30 mL/min heated to 350 °C at a rate of 5 °C/min and then cooled to 60 °C in an N$_2$ atmosphere. Then, samples were treated in a N$_2$O atmosphere at a flux rate of 30 mL/min for 30 min for the oxidation of surface copper atoms according to the following equation:

$$2Cu_{surface} + N_2O \rightarrow Cu_2O_{surface} + N_2. \quad (1)$$

Physically adsorbed N$_2$O molecules were removed by N$_2$ purging at a flow of 30 mL/min for 1 h. Finally, the sample underwent a second reduction with the same parameters as the first one for the calculation of hydrogen consumption required to reduce the surface copper atoms. The dispersion was calculated as the ratio between surface and total metallic copper content, obtained from the hydrogen consumption in the second and the first reduction, respectively, considering the stoichiometry of the reaction.

**Catalytic activity measurements.** To evaluate the catalytic performance of the materials, we performed WGS reactions using 50 mg of fine powder samples diluted with an equal amount of quartz placed in a stainless-steel fixed-bed tube reactor. First, the sample was treated with H$_2$ (reducing atmosphere) at 250 °C for 1 h. Catalytic reactions were then conducted at temperatures ranging from 200 °C to 300 °C, with steps of 25 °C, and the total flow rate of feed was 68 mL/min and 10% vol. CO/N$_2$ mixture, H$_2$O: CO = 3:1 molar ratio. Although the stoichiometric ratio between CO and H$_2$O is 1:1 for the WGS reaction, an excess of water is employed to prevent side reactions and thermodynamically favor the formation of the products.[32] To evaluate the rate of catalyzed reactions, gaseous products were analyzed online in an Agilent 7890 A gas chromatograph equipped with two 6-port valves and a 10-port valve, TCD and FID detectors and three columns, namely, HP pona, Plot-Q, and HP molesiev. Water was condensed and removed before the analysis. All post reactor lines and valves were heated to 180 °C to prevent product condensation. The CO conversion was kept below 10%. As a measure of catalytic efficiency relative to available sites, the turnover frequency (TOF) was calculated using the following expression (Equation 2):

$$TOF(s^{-1}) = \frac{(X_{CO} \times F_{CO} \times Na)}{(m_{cat} \times S_{Cu} \times na)} \quad (2)$$

where $X_{CO}$ represents the CO conversion, $F_{CO}$ is the CO flow (mol.s$^{-1}$), $N_a$ is Avogadro's number (6.023 × 10$^{23}$ atm.mol$^{-1}$), $m_{cat}$ is the weight of catalysts, $S_{cu}$ denotes metallic copper surface area (m$^2$.g$^{-1}$), and $n_a$ designates the number of Cu atoms in a monolayer (1.469 × 10$^{19}$ atoms.m$^{-2}$).[33]

Rates of reaction for the overall WGS reaction on SrTi$_{1-x}$Cu$_x$O$_3$ catalysts with $x$ = 0.15, 0.20 and 0.30 at 200 °C and 1 atm total pressure were measured under differential conditions (conversion < 10%). WGS rates were normalized by the metallic copper area (mol CO m$^{-2}$$_{(Cu)}$ s$^{-1}$)). The reaction rate was calculated using the following expression (Equation 3):

$$rCO = \frac{(CO_{in} - CO_{out})}{W \cdot S_{Cu}}, \quad (3)$$

where rCO is the inlet and outlet flow rate of CO (mol. s$^{-1}$), W is the mass of catalyst (g), and $S_{Cu}$ is the specific surface area of metallic copper (m$^2$. g$^{-1}$). An elementary reaction with first-order kinetics was assumed concerning all species in the WGS reaction,[34] and the apparent activation energy ($E_a$) was calculated by the slope of the fitted lines. Then, the value of the slope corresponds to $E_a/R$, where $R$ is 8.314 J/mol. K.[35]

For the calculation of the equilibrium-limited CO conversion, the equilibrium constant for the WGS reaction, $K_{eq}$, was obtained from Twigg's Catalyst Handbook[36] and related to the concentrations of reagents and products by equation 4:[34,37]

$$K_{eq} = [CO_2][H_2]/[CO][H_2O]. \quad (4)$$

**Computational analysis.** Cu formation energies in the SrTiO$_3$ host lattice and on the surface were calculated by employing density functional theory (DFT) calculations[38] within the projector augmented wave[39–41] formalism as implemented in the VASP code.[42–43] Brillouin zone sampling was carried out by choosing a 7 × 7 × 7 k-points set within a Monkhorst-Pack scheme that generated a grid with a spacing of ~ 0.04 Å$^{-1}$ between k points while the energy cut-off was set to 500 eV. The total energy convergence test was performed by increasing the k point mesh with ~ 0.03 Å$^{-1}$ spacing; it was found that the total energy differs only by 10$^{-6}$ eV/atom, indicating excellent convergence. We used generalized gradient approximations as parameterized by Perdew and Wang for the exchange-correlation functional.[44] An orbital-dependent Hubbard term ($U$)[45] of the effective value of 4 eV was applied to both Ti and Cu atoms to improve the accuracy of DFT calculations. This value of $U$ significantly improves the calculated band edge positions and defect levels in the Ti and Cu oxides.[46–48] The value of 4 eV for SrTiO$_3$ has particularly been shown to produce an accurate band structure.[48,49] Furthermore, the use of a similar $U$ value for Cu offers both accuracy and ease of interpretation of results.[50]

The formation energy ($E_f$) of Cu atoms when substituting Ti (Cu$_{Ti}$) or Sr (Cu$_{Sr}$) and when located interstitially (Cu$_{Int}$) was calculated within a $3a \times 3a \times 3a$ supercell using the conventional procedure[51,52] as described by the following equation:

$$E_f = E_t(SrTiO_3:Cu) + \mu_\alpha - E_t(SrTiO_3) - \mu_{Cu}, \quad (5)$$

Here, $E_t$(SrTiO$_3$:Cu) is the total energy of the SrTiO$_3$ supercell containing the Cu species, and $E_t$(SrTiO$_3$) is the total energy of the pristine SrTiO$_3$ supercell. $\mu_\alpha$ and $\mu_{Cu}$ are the chemical potentials of the removed (Sr or Ti) and added (Cu) elements, respectively. $\Delta\mu$ values were set equal to the total energy of the elemental metallic phase ($\mu_{Metal}$) plus a thermodynamically defined shift ($\Delta\mu$). Based on thermodynamic phase stability considerations, $\Delta\mu$ can be chosen to represent either oxygen-poor or oxygen-rich environments. The detailed derivation of $\Delta\mu$ is given in the Supporting Information. To study the aggregation of a copper species in SrTiO$_3$, the total energies of four configurations of a $3a \times 3b \times 3c$ supercell containing two Cu species with varying Cu–Cu distances ($d_{Cu-Cu}$) were calculated and compared. The relative energy of these configurations ($E_R$) is defined as the difference in the total energy of any given configuration with the total energy of the most stable one.

## RESULTS AND DISCUSSION

**Structural Characterizations.** Figure 1 shows the XRD patterns of the $x = 0.15$, 0.20 and 0.30 samples. In good agreement with our earlier work, the main crystalline phase was identified as a perovskite structure with space group $Pm\bar{3}m$ with peaks corresponding to those of cubic $SrTiO_3$ (PDF 35-0734) present alongside additional minor peaks attributed to the $SrCO_3$ phase (PDF 5-418). As expected, the segregation of the CuO phase was evidenced through the main diffraction peaks of this phase located at 35.6° and 38.7°.[29]

Table S1 lists the lattice parameters of the perovskite phase and the fraction of the segregated CuO and $SrCO_3$ phases, as determined by Rietveld refinement. The lattice parameters for the perovskite phase in the $x = 0.20$ and $x = 0.30$ materials are similar and show little smaller values in comparison to the $x = 0.15$ sample ($a = 3.9171$ Å). Considering the larger ionic radius of $Cu^{2+}$ (0.73 Å) relative to $Ti^{4+}$ (0.61 Å), an expansion in the unit cell, which leads to an increasing in the lattice parameter, might be expected with the substitution of titanium by copper. On the other hand, the valence difference between Cu and Ti can bring about an energetic compensation manifesting in a shift in the position of energy bands and compensation in the unit volume, leading to a contraction of lattice parameter despite the larger size of the substituting Cu ion, as discussed by Janotti et al.[53] The quality of the refinement is indicated by R and $\chi^2$ factors, which are close to the expected values.[54] The amount of the segregated CuO phase was also determined (Table S1) to be equal to 5.10% ($x = 0.15$), 6.4% ($x = 0.20$) and 9.2% ($x = 0.30$). Regarding the $SrCO_3$ phase, the $x = 0.15$ and $x = 0.30$ samples present around 18% while in the sample $x = 0.15$, this value is only 6.5%.

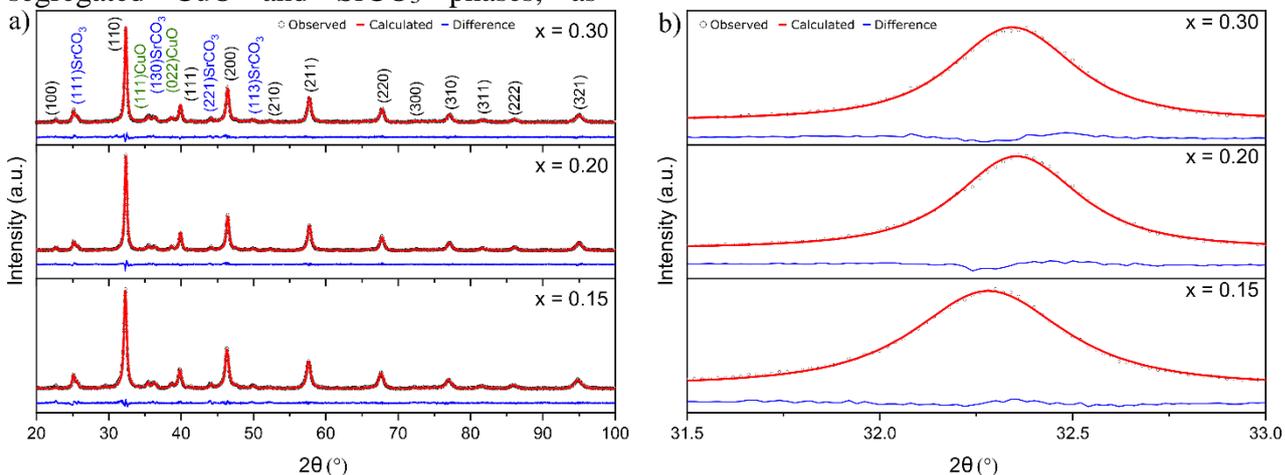

**Figure 1. (a) X-ray diffraction patterns of $SrTi_{1-x}Cu_xO_3$ and (b) The shift in the main $SrTiO_3$ diffraction peak.**

To evaluate the copper dispersion at smaller length scales, we obtained STEM images using a HAADF detector (Figure 2). EDX mapping of Cu was also carried out with STEM to confirm the distribution of copper-rich regions. In the $x = 0.15$ sample, the copper atoms are observed to be well dispersed in the matrix and form very small particles in the CuO phase. In the $x = 0.20$ sample, it is possible to observe the formation of some larger particles in the CuO phase, even more so with high dispersion, whereas in the $x = 0.30$ sample, the aggregation of smaller CuO particles is observed.

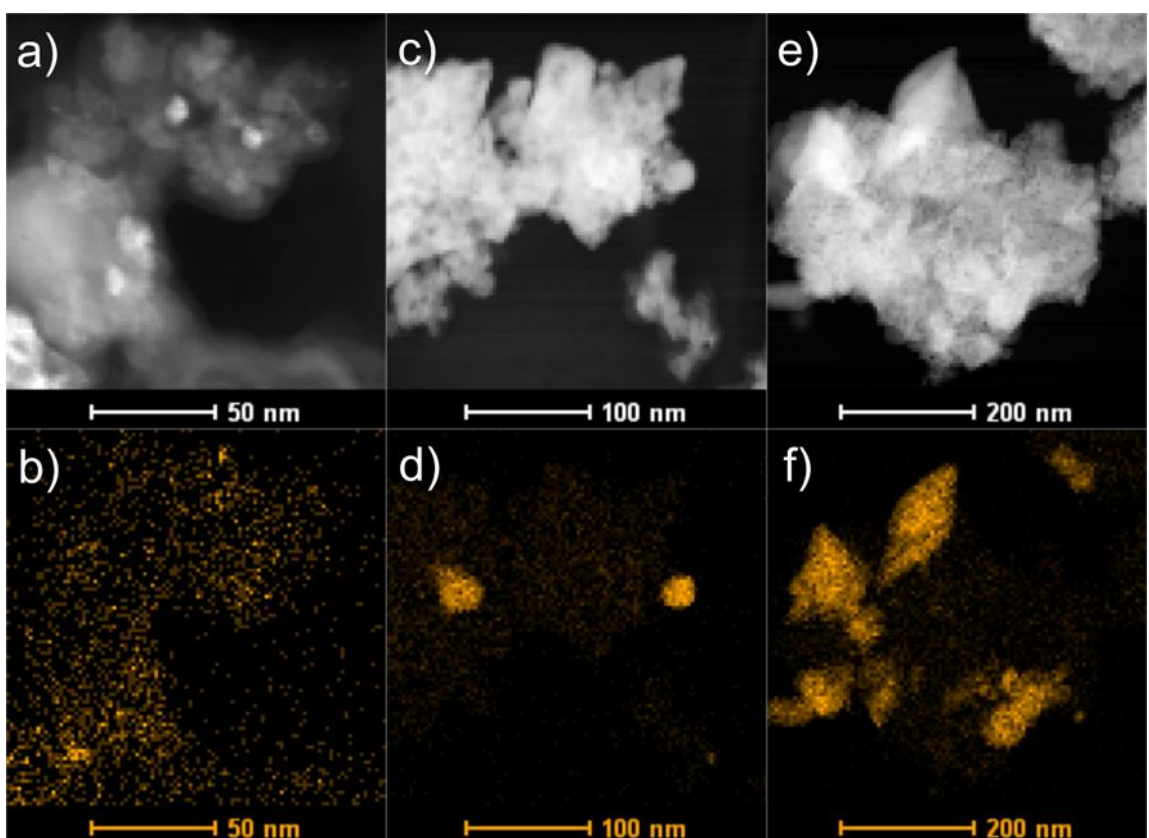

**Figure 2.** STEM HAADF and corresponding EDX mapping images of SrTi$_{1-x}$Cu$_x$O$_3$ with $x$ = 0.15 (a, b), 0.20 (c, d) and 0.30 (e, f) samples.

Additionally, we evaluated the dispersion of metallic particles on the surface of catalysts following the reduction step (N$_2$O-TPD). It was found that the dispersion of metallic copper on the surface of the perovskites decreases with increasing copper content to 11% ± 1 for $x$ = 0.30 catalyst compared to the $x$ = 0.15 and $x$ = 0.20 catalysts that presented dispersions of 15% ± 1 and 16% ± 1, respectively. The decrease of the dispersion of metallic copper is probably due to the increase of CuO particle size or aggregation of smaller CuO particles on the surface of the catalysts with the increasing of copper loading, as observed by STEM HAADF (Figure 2). High-resolution transmission electron microscopy (HRTEM) images (Figure 3) show atomic planes separated at distances of approximately 0.24 nm, corresponding to the (111) plane of the CuO phase (PDF 48-1548).

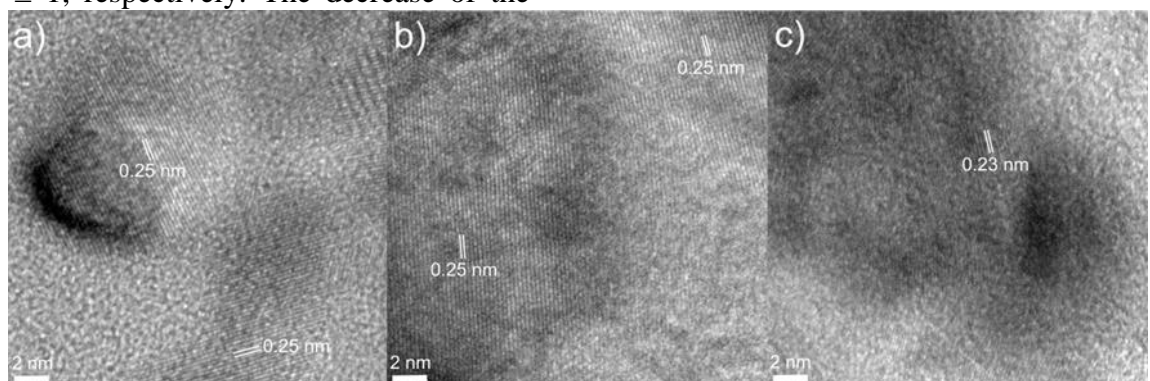

**Figure 3.** HRTEM images of SrTi$_{1-x}$Cu$_x$O$_3$ with $x$ = 0.15 (a), 0.20 (b) and 0.30 (c) samples.

Figure 4 shows TPR profiles acquired from $SrTiO_3$ and $SrTi_{1-x}Cu_xO_3$ with $x = 0.15$, 0.20, and 0.30 during heating and under the reducing atmosphere. Catalysts showed a main peak at approximately 200 °C, which can be ascribed to the reduction of surface CuO to metallic copper.[17] This peak is more significant for $x = 0.30$ as the result of higher levels of the secondary copper oxide phase in this material. For further investigation of the TPR results, all profiles were deconvoluted by using a Gaussian function (Figure S1). The profile of the catalyst with $x = 0.20$ was deconvoluted into 3 peaks (α, β, and γ in our terminology), suggesting the presence of different Cu species and/or Cu particle sizes. On the other hand, the $x = 0.15$ and $x = 0.30$ catalysts profiles showed only a broad peak related to the copper reduction.

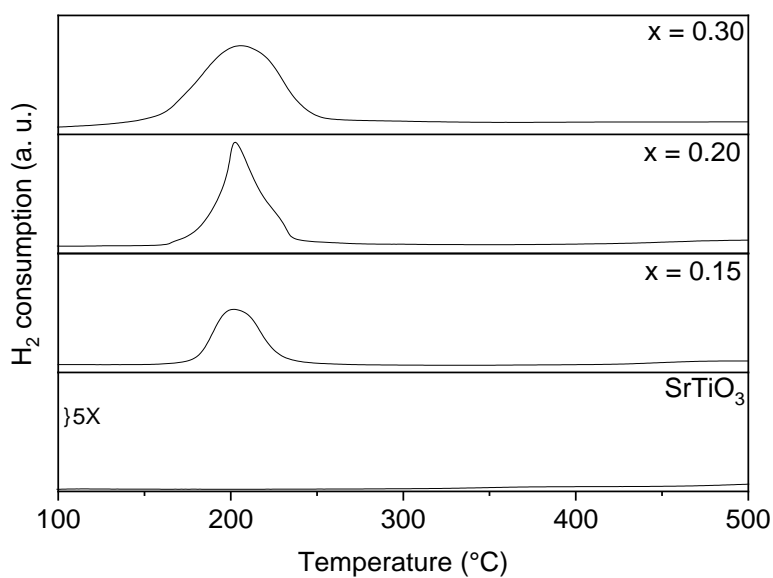

**Figure 4.** TPR profiles of $SrTiO_3$ ($x = 0.00$) and $SrTi_{1-x}Cu_xO_3$ with $x = 0.15$, 0.20 and 0.30.

**Catalytic Performance.** Figure 5 shows the Arrhenius-type plot of turnover frequency as a function of reaction temperature over $SrTi_{1-x}Cu_xO_3$ materials with $x = 0.15$, 0.20, and 0.30. The turnover frequency was strongly dependent on the temperature of the reaction for all samples. TOF values obtained varied from 0.05 to 1.33 ($s^{-1}$), values of the same order of magnitude as those shown in the literature for Cu-based catalysts,[11,18,55–59] (Table S2 of the Supporting Information). It is essential to highlight that the $x = 0.20$ catalyst showed the highest TOF for all temperatures evaluated when compared with $x = 0.15$ and 0.30. Probably, the TOF achieved for the catalyst with $x = 0.20$ is related to the different Cu species and or different particle sizes of Cu species when compared with the $x = 0.15$ and 0.30 catalysts, as presented in Figure S1.

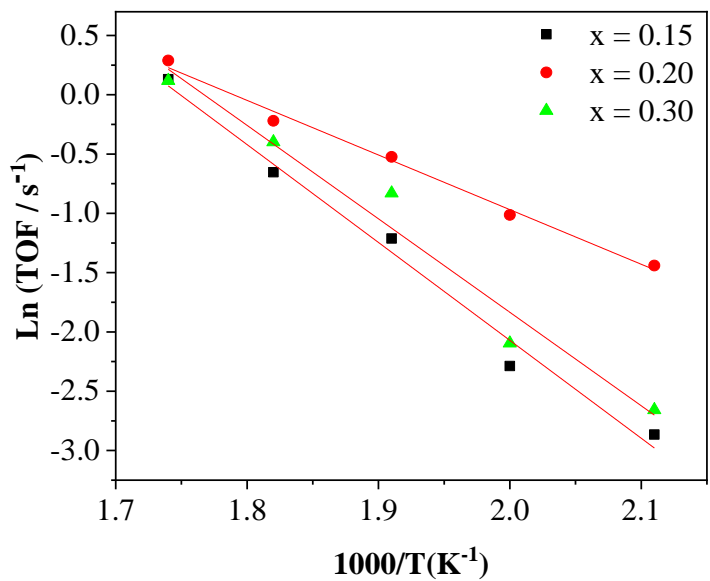

**Figure 5.** Arrhenius-type plots of turnover frequency for the WGS reaction over $SrTi_{1-x}Cu_xO_3$ catalysts with $x$ = 0.15, 0.20, and 0.30.

Figure 6 shows the WGS rate relative to the available metallic copper surface area at 200 °C and alongside the apparent activation energy ($E_a$) for $SrTi_{1-x}Cu_xO_3$ catalysts. The copper metallic areas of these catalysts were 21, 23, and 24 $m^2 \cdot g^{-1}$ for $x$ = 0.15, 0.20 and 0.30, respectively. As demonstrated, increasing the CuO crystalline phase higher than 6.4% ($x$ = 0.20 sample) on $SrTi_{1-x}Cu_xO_3$ catalysts did not yield a monotonic increase in the TOF or WGS rate.

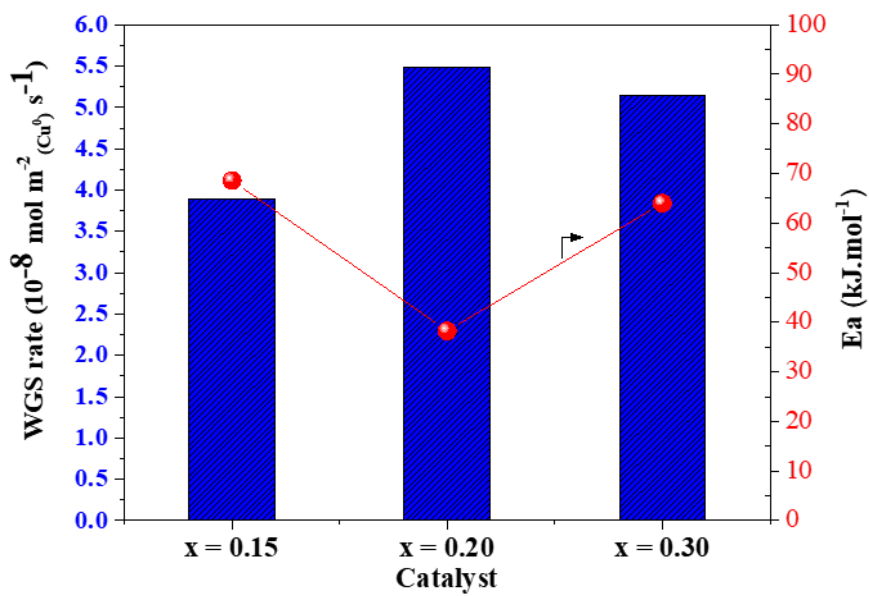

**Figure 6.** Rates of reaction for the overall WGS reaction, normalized per unit of copper surface area at 200 °C and apparent activation energy ($E_a$) for $SrTi_{1-x}Cu_xO_3$ catalysts with $x$ = 0.15, 0.20, 0.30.

The apparent activation energy ($E$a) was calculated from the slopes of the lines in the Arrhenius plots of Figure 5. The apparent activation energies reached 68.6, 38.2 and 64.0 kJ mol$^{-1}$ for SrTi$_{1-x}$Cu$_x$O$_3$ with $x$ = 0.15, 0.20 and 0.30, respectively. The apparent activation energy achieved for SrTi$_{1-x}$Cu$_x$O$_3$ with $x$ = 0.20 catalyst was close to the values of the best WGS catalysts presented in the literature (ca. 30–37 kJ/mol),[34,59–66] (Supporting Information, Table S2).

It is worth emphasizing that the $x$ = 0.20 material showed an apparent $E$a almost 50% lower than that of the other samples that were evaluated, and this apparent $E$a is close to the reported values for Cu$_{0.2}$Ce$_{0.8}$O$_{2-y}$ WGS catalysts ($E$a = 34 kJ/mol) under conditions of 473–623 K and CO/H$_2$O = 1/3.[34] The enhanced catalytic activity of the presently studied systems is further evident from a comparison with reported activation energies for WGS on pristine Cu, reported as 63.6, 42.0 and 71 kJ/mol for Cu(100), Cu(110) and Cu(111) surfaces, respectively,[68] In general, Cu supported on oxide supports is reported to exhibit lower WGS activation energies than pristine copper, with $E$a values between ca. 30 and 80 kJ mol$^{-1}$,[18,63,69]

Reaction rates are somewhat more challenging to compare relative to $E$a values owing to the divergence in feed rates, pressures, and temperatures. The reaction rates in the present work are within the proximity of normalized reaction rates for various other metal-promoted WGS catalysts,[63] (Tables S3 and S4 of the Supporting Information). However, a meaningful quantitative comparison is rendered moot by the divergence in reaction conditions.[63]

The most promising WGS catalyst candidates studied to date, as mentioned before, are ceria-supported systems. The results found here do not present a significant improvement over CeO$_2$-supported copper. However, as a first investigation into the use of perovskites in WGS catalysis, the outcomes demonstrate the viability of such systems and their merit toward WGS applications.

Additional tests at higher CO conversion and to evaluate the thermal stability are carried out in Figure 7. Figure 7a shows the WGS reaction results for the SrTi$_{1-x}$Cu$_x$O$_3$ sample with $x$ = 0.03, 0.15, 0.20 and 0.30 until reaching the steady-state. For all three samples, the CO conversion increases substantially with increasing temperature, reaching 5, 44, 74, and 46% for $x$ = 0.03, 0.15, 0.20, and 0.30, respectively, at a higher temperature (350 °C). The low activity for the sample with $x$ = 0.03 indicates that the pristine SrTiO$_3$ compound is inert for the WGS reaction. The activity for any fixed temperature is highest for $x$ = 0.20, as was discussed before. Figure 7b further shows the stability test for $x$ = 0.20, which presented the best activity at 350 °C for 10 h. The catalyst shows good stability until 6 h of reaction and a decrease of approximately 10% in CO conversion until 10 h. Sintering of copper is usually a cause for catalyst deactivation.[70]

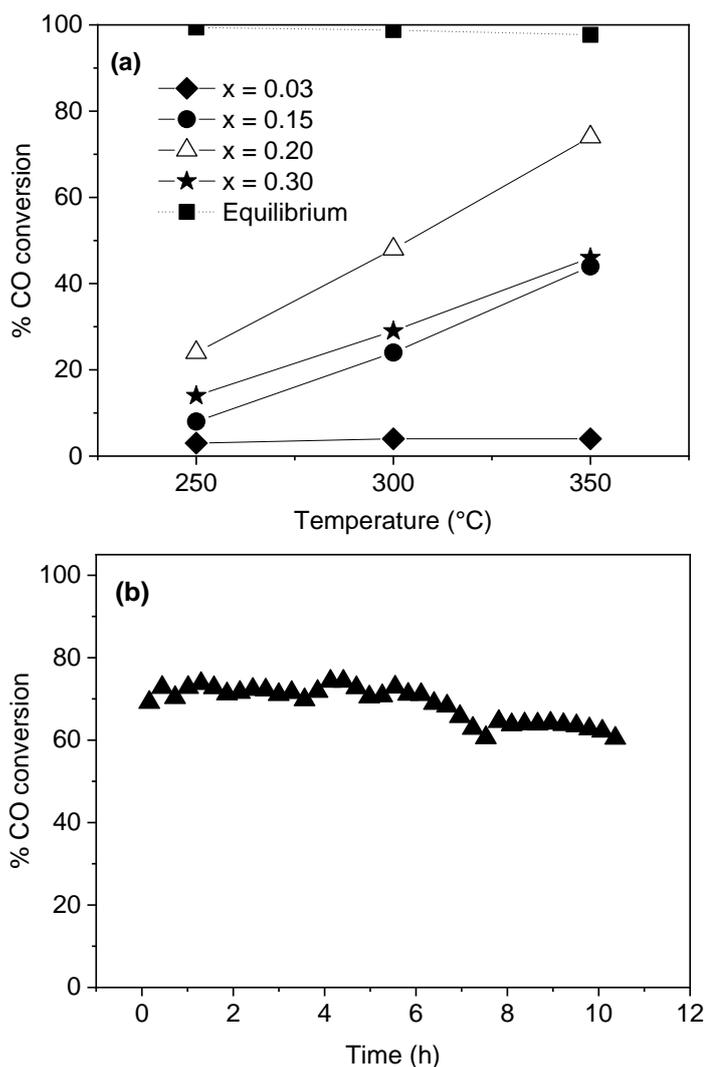

**Figure 7.** CO conversion on the SrTi$_{1-x}$Cu$_x$O$_3$ catalyst with $x$ = 0.15, 0.20 and 0.30 (a) and thermal stability test for SrTi$_{1-x}$Cu$_x$O$_3$ catalyst with $x$ = 0.20 at 350 °C (b). Conditions: 180 mg of catalysts under a gas atmosphere of 5 vol%. CO/N$_2$ at a flux rate of 100 mL/min and CO:H$_2$O = 1:3 molar ratio.

**Structural Characterization in Operando Conditions.** To gain insights into the oxidation Cu state and the behavior of perovskite during the WGS reaction, in situ X-ray diffraction and X-ray absorption experiments were also carried out at 350 °C. Figures 8a and 8b compare the diffraction patterns of samples in the 35–45° 2$\theta$ range at room temperature before reaction and after reduction and 1 h of WGS reaction at 350 °C. As can be observed, the diffraction patterns showed the presence of CuO, Cu, SrCO$_3$, and SrTiO$_3$ phases before and after 1 h of WGS reaction, indicating that the copper is still metallic and the obtained phase assemblage is stable throughout the reaction for the examined composition range. Calculations of crystallite size using the Scherrer equation show that regardless of composition, the crystalline size of the Cu phase is on the order of 40 nm.

**Figure 8.** XRD of SrTi$_{1-x}$Cu$_x$O$_3$ samples. $x = 0.15$ sample: (a) at room temperature before reduction and reaction and (b) after reduction and 1 h of WGS reaction at 350 °C; $x = 0.20$ sample: (c) at room temperature before reduction and reaction and (d) after reduction and 1 h of WGS reaction at 350 °C; $x = 0.30$ sample: (e) at room temperature before reduction and reaction and (f) after reduction and 1 h of WGS reaction at 350 °C.

Figures 9a and 9b show the XANES spectra at Cu of the SrTi$_{1-x}$Cu$_x$O$_3$ samples after reduction and with the spectra collected during the WGS reaction at 350 °C. Similar to the XANES spectra of the Cu phase, all Cu atoms belonging to the CuO phase were reduced, and if some Cu species presented on the matrix of the SrTiO$_3$ phase were not reduced, they could not be detected, probably due to the small number of Cu atoms inside of the SrTiO$_3$ phase (less than 6 mol% according to our previous work.[29]). These data are in good agreement with the in situ X-ray diffraction (Figure 8b). Figure 9b shows the Ti K-edge XANES spectra of the $x = 0.15$ sample collected at room temperature before reaction and after reduction, and during the WGS reaction conditions. As seen in Figure 9b, no changes in the Ti K-edge XANES spectrum are observed between measurements made under ambient conditions and during the WGS reaction, allowing us to conclude that only the copper atoms were reduced while the oxidation state and the local structure of titanium atoms remained stable during the WGS reaction. This result is in line with the XRD data, which show that the long-range order structure of the SrTiO$_3$ perovskite phase remains stable under the conditions of the WGS reaction.

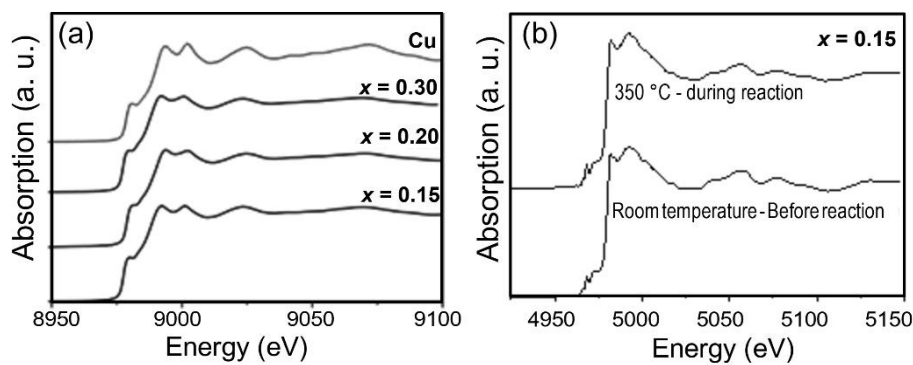

**Figure 9.** (a) In situ Cu K-edge XANES spectra of SrTi$_{1-x}$Cu$_x$O$_3$ samples after reduction and 1 h of WGS reaction at 350 °C. (b) XANES spectra at the Ti edge of sample $x = 0.15$ at room temperature and after reduction and 1 h of WGS reaction at 350 °C.

**Cu Aggregation Tendencies from DFT Calculations**. Computational analysis based on DFT was conducted to glean new insights into the behavior of Cu within the SrTiO$_3$ matrix that initiates the observed Cu agglomerates' growth. The formation energy ($E_f$) of incorporating Cu atoms in the SrTiO$_3$ host matrix was calculated as a function of O's chemical potential (Figure 10a) and is shown in Figure 10b. Under the O-rich environment, Point *B*, Cu$_{Ti}$ species with an $E_f$ of 0.481 eV were comfortably more stable than Cu$_{Sr}$ ($E_f$ = 1.706 eV) and Cu$_{Int}$ ($E_f$ = 4.945 eV). The sequence of stabilization was, however, reversed under the O-poor environment, with Cu$_{Int}$ being the most stable site with an $E_f$ of 4.756 eV. Consequently, since our samples were treated under an oxygen environment at 500 °C during the synthesis, we can assume that all Cu atoms initially substituted a Ti ion. The preference of Cu substituting Ti under the O-rich environment has also been observed for samples prepared by solid-state reaction[25] and has been predicted in DFT calculations based on the GGA method.[27] In all these cases, the oxidation of Cu dopant can be inferred from the DFT site projected spin population on Cu, which is 0, pertaining to d$^{10}$ electronic configuration, for Cu$_{Int}$ and Cu$_{Sr}$, and 0.952 *e*, pertaining to d$^9$ electronic configuration, for Cu$_{Ti}$, indicating that Cu$_{Int}$ and Cu$_{Sr}$ have an oxidation state of +1, while Cu$_{Ti}$ has an oxidation state of +2.

In the next stage, the aggregation of the Cu$_{Ti}$ species was studied to examine the propensity toward CuO secondary phase formation. This was achieved by varying the Cu–Cu distance in a supercell containing two Cu$_{Ti}$ atoms and monitoring the total energy. The configurations are shown in Figure 10c. In each configuration, one Cu atom was placed at site **o**, while another Cu atom was placed at sites **a**, **b**, **c**, and **d**, for which the total relative energy ($E_R$) is given in Figure 10d. The total energy of the supercells containing two Cu$_{Ti}$ varies only marginally as a function of $d_{Cu–Cu}$. For instance, the shortest possible $d_{Cu–Cu}$ for Cu$_{Ti}$ was 3.95 Å, for which the total energy was only 0.064 eV higher than that of the most stable configuration ($d_{Cu–Cu}$ = 8.88 Å). Given the weak relationship between the total energy and $d_{Cu–Cu}$ for Cu$_{Ti}$, the distribution of Cu$_{Ti}$ in the SrTiO$_3$ host lattice is predicted to be homogeneous for dilute Cu concentrations.

In contrast, for higher Cu concentrations, the formation of secondary CuO or CuO$_2$ phases or ternary oxides is likely dominated by random phase percolation.[71] The same trend was also predicted for Cu$_{Ti}$ at the SrTiO$_3$ (100) surface, as shown in Figure S2. Of particular note, the absence of a strong thermodynamic driving force for either aggregation or dispersion of Cu$_{Ti}$ species suggests that the local structure of copper species in SrTiO$_3$ prepared under O-rich conditions is likely to be sensitive to the applied synthesis protocols. The synthesis methodology used here is conducive to steric entrapment of cations and would be expected to yield lower levels of phase segregation for smaller Cu concentrations, as seen in the EDX mapping relative to other synthesis methods that may be prone to inhomogeneous crystallization during formation.[72]

Consequently, the agglomeration observed in the STEM images is driven by random percolation growth along with the higher likelihood of Cu$_{Ti}$ ions occupying nearest neighbor sites as Cu concentration increases. For the sample with *x* = 0.15, the low turnover frequency can be attributed to the smaller number of active sites. For *x* = 0.20, a higher Cu$_{Ti}$ concentration increases the active sites and, in turn, the turnover frequency. For *x* = 0.30, the ever-larger agglomeration of Cu$_{Ti}$, and therefore the resultant secondary CuO areas on the surface, as seen in the EDX mapping of Cu (Figure 2), eventually reduces the exposed active sites (metallic Cu) available for the catalytic reaction, reducing the turnover frequency.

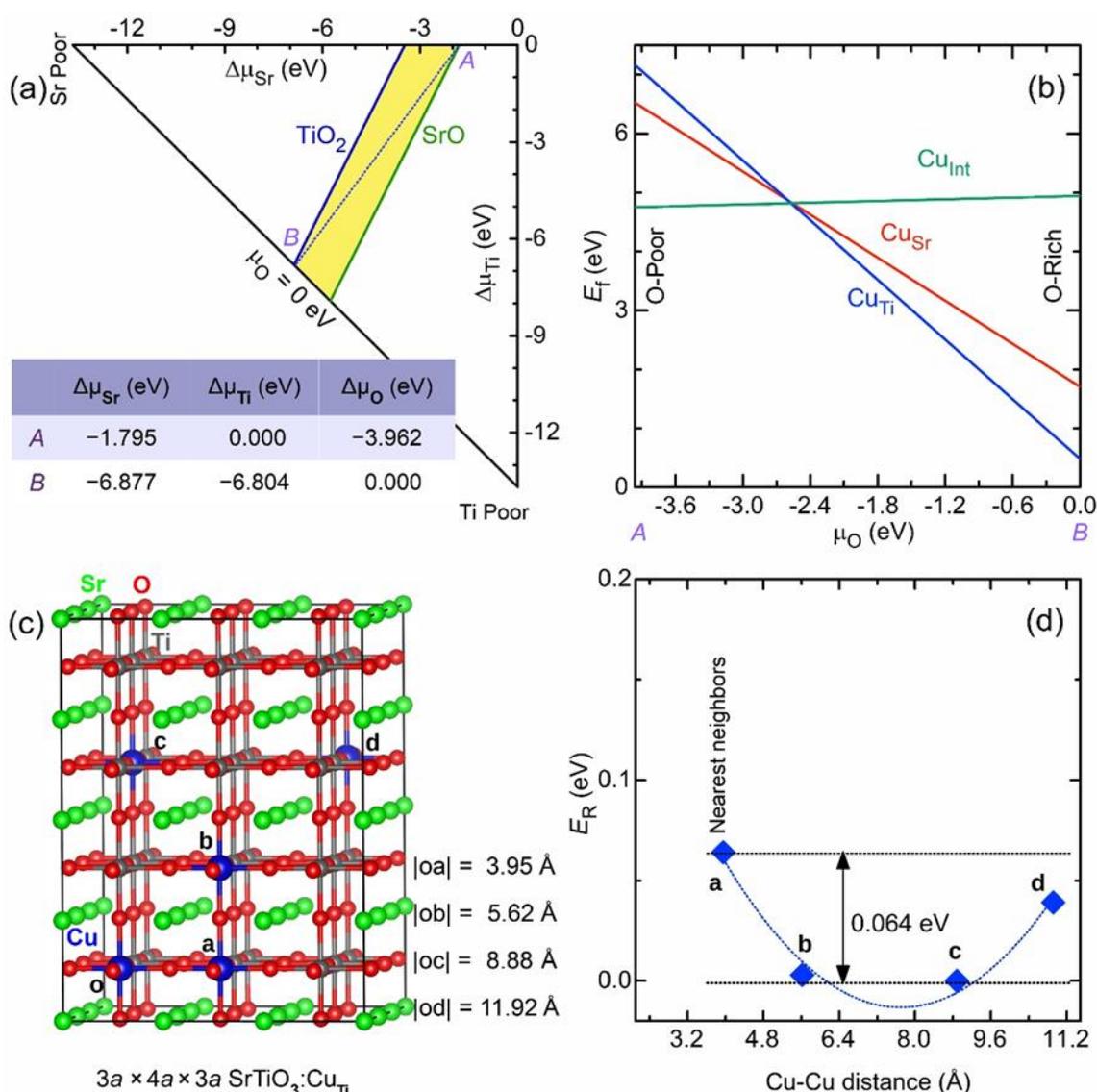

**Figure 10.** (a) The phase diagram for cubic SrTiO$_3$ at zero Kelvin. The shaded area represents the chemical stability range of SrTiO$_3$. The chemical potentials at *A* and *B* are referred to as oxygen-poor and oxygen-rich, respectively. (b) The formation energy ($E_f$) of Cu atoms in the SrTiO$_3$ host lattice under the O-poor environment (left side) and O-rich environment (right site). (c) Configurations containing two Cu atoms that were used to study the Cu$_{Ti}$ aggregation tendency. In each configuration, one Cu atom is located at site o, while the second Cu ion is located at either a, b, c, or d. (d) The relative total energy ($E_R$) for configurations with two Cu atoms as a function of Cu–Cu distance. $E_R$ is the total energy value given relative to the most stable configuration.

Furthermore, the DFT formation enthalpy of SrTiO$_3$, defined as $\Delta H^f(\text{SrTiO}_3) = E_t(\text{SrTiO}_3) - E_t(\text{Sr}) - E_t(\text{Ti}) - 1.5E_t(\text{O}_2)$, was found to be $-13.681$ eV per formula unit (f.u.), while the formation enthalpy of CuO, $\Delta H^f(\text{CuO}) = E_t(\text{CuO}) - E_t(\text{Cu}) - 0.5E_t(\text{O}_2)$, was calculated to be merely $-0.189$ eV/f.u. The significantly smaller (negative) formation enthalpy of CuO with respect to that of SrTiO$_3$ ensures that in the catalysis reduction step, CuO is preferentially reduced rather than SrTiO$_3$, leaving exposed catalytically active metallic Cu sites on the SrTiO$_3$ surface. The presence of metallic Cu on SrTiO$_3$'s surface is well-known to facilitate the CO adsorption, which is the first step of the WGS catalytic reactions,[73,74] as the affinity of CO to pristine SrTiO$_3$ is rather weak.[75]

Figure 11 shows the electronic density of states (DOS) of the Cu-modified SrTiO$_3$. The valence band is mainly composed of O 2p states, while the conduction band is primarily composed of Ti 3d states. With Cu$_{Ti}$ species, the octahedral coordination of the original Ti site splits the Cu 3d states, therefore creating an empty band of the $e_g$ states in the middle of the fundamental bandgap, resulting in a deep copper band that manifests as catalytically active levels in SrTi$_{1-x}$Cu$_x$O$_3$. These copper states can offer chemically labile empty orbitals for accepting electrons from CO's 5σ molecular orbitals, as shown in the inset, and facilitate the chemisorption of CO molecules on the Cu-rich areas of the surface.[76,77]

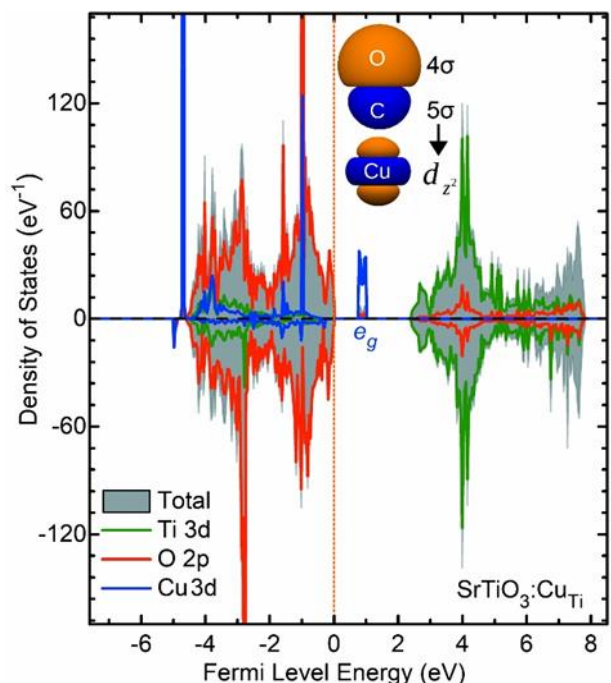

Figure 11. The total and partial density of states of SrTiO$_3$:Cu$_{Ti}$. Copper's states are magnified by a factor of 10 for clarity. The inset schematically shows how the labile empty Cu states can facilitate the adsorption of the CO molecule produced with the Orbitals Viewer package.[78]

## CONCLUSIONS

Cu-modified SrTiO$_3$ (SrTi$_{1-x}$Cu$_x$O$_3$) was studied as a novel material for water-gas shift (WGS) catalysis. Among the samples prepared here, with different copper contents, there are no significant differences in short- and long-range order and oxidation state, both in ambient and in reaction conditions. Using DFT calculations, it was demonstrated that Cu species are likely to occupy Ti sites in the SrTiO$_3$ host matrix under the synthesis conditions applied here. Experimental results show a combination of copper in the host lattice in addition to a precipitating secondary copper oxide phase on the surface, which is readily reduced to a metallic state under WGS reaction conditions. SrTi$_{1-x}$Cu$_x$O$_3$ materials prepared with $x = 0.20$ exhibited higher WGS rates relative to the lower apparent activation energy relative to materials of lower and higher Cu content. The enhanced catalytic performance for the $x = 0.20$ catalyst is related to the optimized dispersion of Cu$^0$ sites on the perovskite lattice, which promotes WGS reactions. Catalytic activity was found to compare favorably with earlier studies of metal-loaded oxide WGS catalysts. The results here serve to establish the appropriateness of Cu-substituted strontium titanate materials for use as water gas shift catalysts and further highlight the central role of available metallic sites in governing the kinetics of catalyzed reactions.

## ASSOCIATED CONTENT

Supporting Information

Experimental evaluation of the amount of metallic Cu, lattice parameters of $SrTi_{1-x}Cu_xO_3$, activation energies, turnover frequencies, and reaction rates compared with the literature, permissible chemical potential ($\Delta\mu$) values for density functional calculations, calculated segregation tendency among Cu atom on the surface of $SrTiO_3$, calculated segregation tendency of Cu under O-poor environment

## ACKNOWLEDGMENTS

The authors are grateful for the financial support provided by São Paulo Research Foundation – FAPESP (grant 2013/07296-2, 2017/08293-8 and 2015/06246-7) and National Council for Scientific and Technological Development – CNPq (grants 304498/2013-0, 304883/2016-6 and 140631/2013-5). Our thanks are also extended to the Brazilian Synchrotron Light Laboratory (LNLS) for the use of its XAFS2 and XPD beamline experimental facilities (proposal numbers 15960 and 17001). The authors are also grateful to Alessandra F. Lucredio for TPR measurements and Vitor A. S. Mendes for the use of the transmission electron microscope. Computational resources were provided by supercomputers at the Center for Computational Sciences at the University of Tsukuba, Japan.